\documentclass{article}

\usepackage{amsfonts, amsmath, amssymb, amsbsy}
\usepackage[]{graphicx}
\usepackage{jwdiplbib}

\begin{document}
\bibliographystyle{jwaabib}

\begin{center}
\textbf{\LARGE Structural variability of 3C\,111 on parsec scales}
\end{center}

\begin{center}
\textbf{C. Grossberger$^1$, M. Kadler$^{2,1}$, J. Wilms$^1$, C. M\"uller$^1$, T. Beuchert$^1$, E. Ros$^{3,4}$, R. Ojha$^5$, M. Aller$^6$, H. Aller$^6$, E. Angelakis$^4$, L. Fuhrmann$^4$, I. Nestoras$^4$, R. Schmidt$^4$, J.A. Zensus$^4$, T.P. Krichbaum$^4$, H. Ungerechts$^7$, A. Sievers$^7$, and D. Riquelme$^7$}
\end{center}

\begin{center}
{\it
\noindent $^1$Dr. Remeis Sternwarte \& ECAP, Universit\"at Erlangen-N\"urnberg, Sternwartstr. 7, 96049 Bamberg, Germany \\
$^2$Institut f\"ur Theoretische Physik und Astrophysik, Universit\"at W\"urzburg, Am Hubland, 97074 W\"urzburg, Germany \\
$^3$Departament d'Astronomia i Astrof\'isica, Universitat de Val\`encia, 46100 Burjassot, Val\`encia, Spain\\
$^4$Max-Planck-Institut f\"ur Radioastronomie, Auf dem H\"ugel 69, 53121 Bonn, Germany \\
$^5$Goddard Space Flight Center, NASA, Greenbelt, 8800 Greenbelt Rd., Greenbelt, MD, 20771, USA \\
$^6$Department of Astronomy, University of Michigan, Ann Arbor, MI, 48109-1042, USA \\
$^7$Instituto de Radio Astronom\'ia Milim\'{e}trica, Avenida Divina Pastora 7, Local 20, 18012, Granada, Spain\\
 }
\end{center}

\begin{abstract}
We discuss the parsec-scale structural variability of the extragalactic jet 3C\,111 related to a major radio flux density outburst in 2007. The data analyzed were taken within the scope of the MOJAVE, UMRAO, and F-GAMMA programs, which monitor a large sample of the radio brightest compact extragalactic jets with the VLBA, the University of Michigan 26\,m, the Effelsberg 100\,m, and the IRAM 30\,m radio telescopes. The analysis of the VLBA data is performed by fitting Gaussian model components in the visibility domain. We associate the ejection of bright features in the radio jet with a major flux-density outburst in 2007. The evolution of these features suggests the formation of a leading component and multiple trailing components.

\end{abstract}

\section{Introduction}
Jets of active galactic nuclei (AGN) are among the most fascinating objects in the universe. From the time when the term ``jet" was first introduced by \cite*{1954ApJ...119..215B} until today it is still unclear how these jets are created and formed.
A prime source to gain insight into the physics of extragalactic jets is the broad-line radio galaxy 3C\,111 (PKS B0415+379) at $z$ = 0.049\footnote{Assuming $H_{o} = 71\,\mathrm{km\,s^{-1}}\,\mathrm{Mpc^{-1}} , \Omega_{\lambda} = 0.73$ and $\Omega_{m} = 0.27$ $(1\,\mathrm{mas} = 0.95\,\mathrm{pc} $; $1\,\mathrm{mas\,yr^{-1}} = 3.1\,\mathrm{c}$)}. The object can be described with a classical FR II morphology (\cite{1974MNRAS.167P..31F}) exhibiting two radio lobes with hot spots and a single-sided jet (\cite{1984ApJ...279...60L}). Untypical for radio galaxies, a small inclination angle of only  $18\,^{\circ}$ to our line of sight has been determined on parsec scales (\cite{2005AJ....130.1418J}). Moreover, 3C\,111 has a blazar-like spectral energy distribution (SED) (\cite{2008ApJ...688..852H}) and shows one of the brightest radio cores in the mm-cm wavelength regime of all FRII radio galaxies. Superluminal motion was detected in this radio galaxy by \cite*{1987A&A...176..171G} and \cite*{1988IAUS..129..105P} making this source one of the first radio galaxies to exhibit this effect. The EGRET source 3EG J0416+3650 has been associated with 3C\,111 (\cite{2005A&A...430..107S}, \cite{2008ApJ...688..852H}) and $\mathrm{\gamma}$-ray emission from 3C\,111 has been confirmed by \textit{Fermi/LAT} (\cite{2010ApJ...715..429A}, \cite{2010ApJS..188..405A}, \cite{2010ApJ...720..912A}). A major flux-density outburst in 1997 was investigated by \cite*{2008ApJ...680..867K} with 10 years of radio monitoring data (1995--2005). In addition \cite*{2011ApJ...734...43C} and \cite*{2011arXiv1108.6095T} report on a possible connection between the accretion disk and the jet of 3C\,111.\\
In this paper a new major flux density outburst from 2007 and the associated jet kinematics will be discussed with data from the Very Large Baseline Array (VLBA\footnote{The National Radio Astronomy Observatory is a facility of the National Science Foundation operated under cooperative agreement by Associated Universities, Inc.}), the University of Michigan Radio Astronomy Observatory (UMRAO\footnote{UMRAO  has been supported by a series of grants from the NSF and NASA and by funds from the University of Michigan}) and the F-GAMMA program.

\section{Data Analysis}
The broad-line radio galaxy 3C\,111 has been part of the VLBA 2cm Survey program since its start in 1995 (\cite{1998AJ....115.1295K}) and its successor MOJAVE (Monitoring Of Jets in Active galactic nuclei with VLBA Experiments, \cite{2009AJ....137.3718L}) in 2002. Twenty-four epochs of data have been taken from 2006 to 2010 within the MOJAVE program of this source. Phase and amplitude self-calibration as well as hybrid mapping by deconvolution techniques were performed as described by \cite*{1998AJ....115.1295K}. Utilizing the program \textsc{DIFMAP} (\cite{1997ASPC..125...77S}), two-dimensional Gaussian components have been fitted in the (\textit{u,v})-plane to the fully calibrated data of each epoch. We refer to the inner $\sim0.5$\,mas as the ``core'' region, which can usually be modeled with two Gaussian components. All models have been aligned by assuming the westernmost components to be stationary and all component positions are measured with respect to it. Conservative errors of 15\% are assumed for the flux densities of the model-fit components accounting for absolute calibration uncertainties and formal model-fitting uncertainties (\cite{2002ApJ...568...99H}). Within the UMRAO radio-flux-density monitoring program (\cite{2003ASPC..300..159A}), more than three decades of single-dish flux-density data have been collected for 3C\,111 at 4.8\,GHz, 8.0\,GHz, and 14.5\,GHz\footnote{In this work, we consider only the 4.8\,GHz UMRAO data. A multi-frequency long-term analysis of the light curve will be presented elsewhere (Grossberger et al., in prep.).}. In addition, 3C\,111 is observed monthly by the F-GAMMA program (\cite{2007AIPC..921..249F}, \cite{2008arXiv0809.3912A}) at multiple frequencies throughout the cm- and mm-bands since 2007.

\section{Results}
\subsection{Lightcurves}
\begin{figure*}
\includegraphics[width=0.45\textwidth]{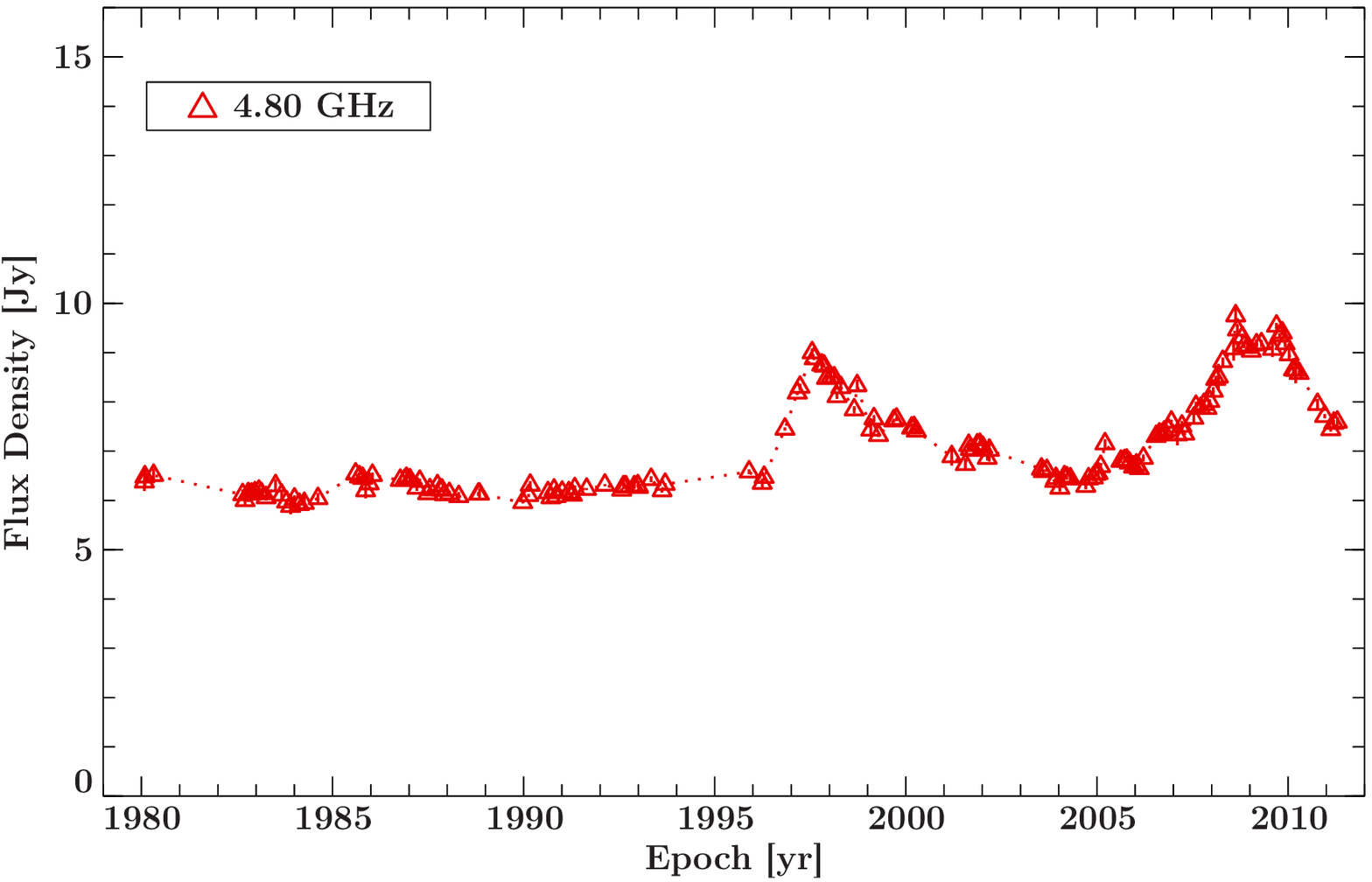}
\includegraphics[width=0.45\textwidth]{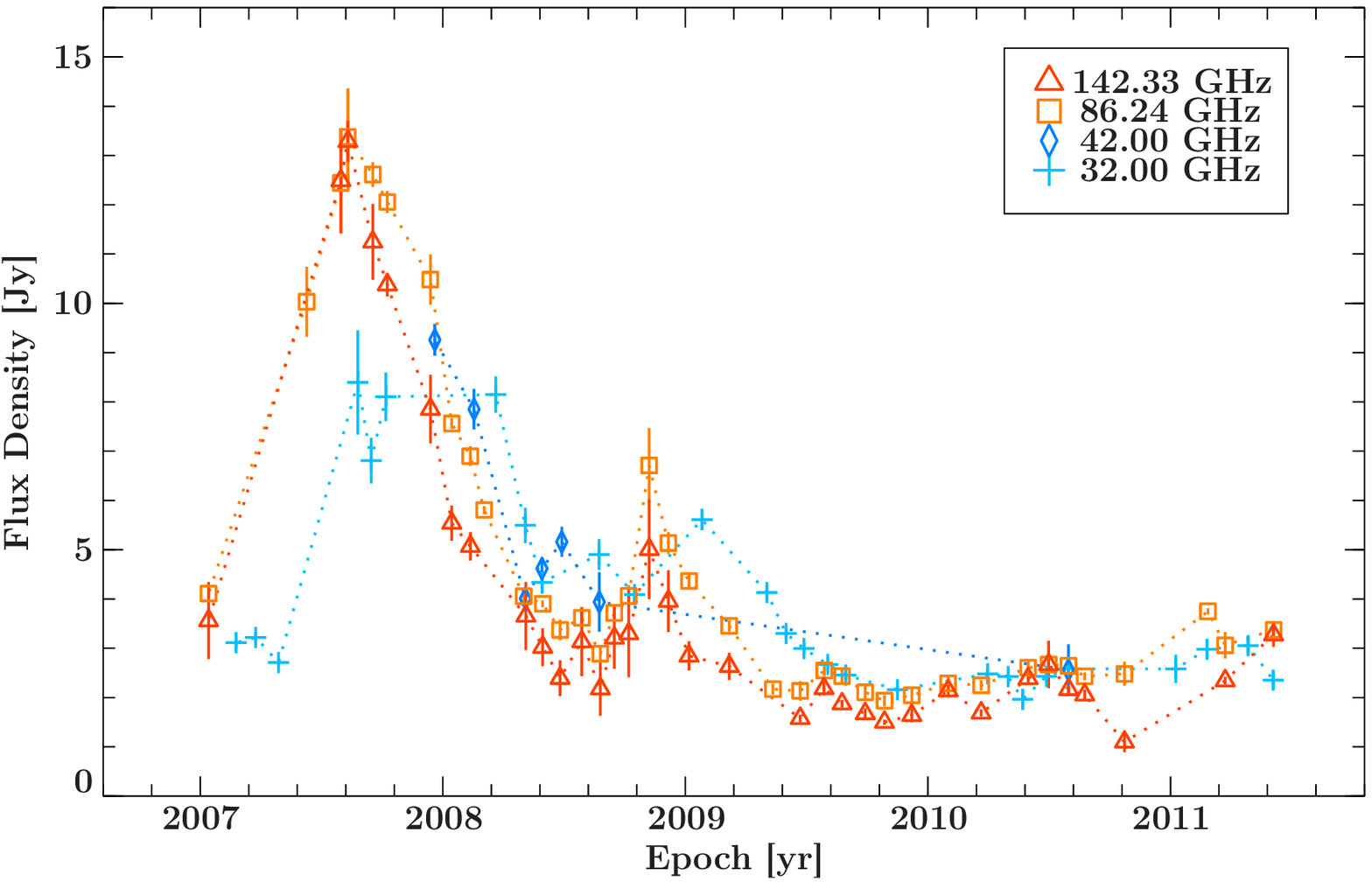}
\caption{Long-term radio lightcurve of 3C\,111 obtained by the UMRAO at 4.8\,GHz (left). Short-term radio lightcurves (right) at 32.0\,GHz, 42.0\,GHz, 86.24\,GHz and 142.33\,GHz obtained by the F-GAMMA program.}
\label{lc_umrao}
\end{figure*}

Figure \ref{lc_umrao} (left) shows the long-term radio lightcurve of 3C\,111 at 4.8\,GHz. Since the start of the measurements the source has been in a low activity state for almost two decades with only minor activity. The source showed a major outburst in 1996/1997 (\cite{2008ApJ...680..867K}). Starting in 2004, minor outbursts are observed and the flux density level increases. A major flux density outburst starts in early 2007 at high frequencies (see Figure \ref{lc_umrao}, right), peaking $\mathrm{\sim2007.6}$ and is subsequently seen at lower frequencies. A secondary outburst starts mid 2008 at high frequencies. The overall flux density level at 4.8\,GHz has been decreasing since the major outburst.

\subsection{VLBA Data Overview}

\begin{figure}
\begin{minipage}[t]{0.305\textwidth}
\vspace{0pt}
\includegraphics[bb = 50 2650 425 3940, clip, width=\textwidth]{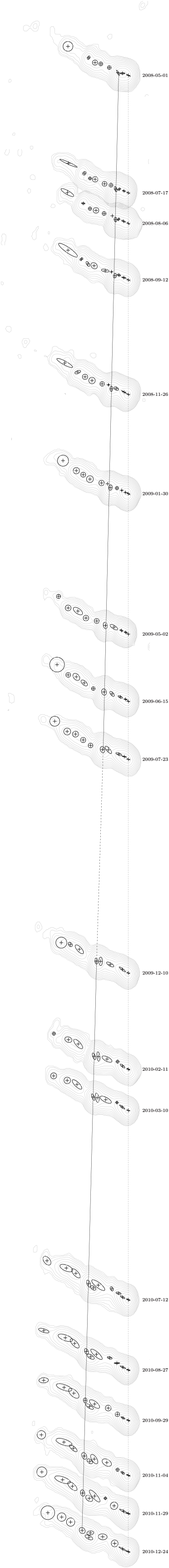}
\end{minipage}
\begin{minipage}[t]{0.305\textwidth}
\vspace{0pt}
\includegraphics[bb = 50 1400 425 2650, clip, width=\textwidth]{eps/test}
\end{minipage}
\begin{minipage}[t]{0.305\textwidth}
\vspace{0pt}
\includegraphics[bb = 50 0 425 1400, clip, width=\textwidth]{eps/test}
\end{minipage}
\caption{Naturally weighted \textsc{clean} images of 3C\,111 from 2008 to 2010. The minimum contours are set to 5 sigma of the rms noise and increase logarithmically by a factor of 2. All fitted model components are indicated with a circle or ellipse (size of the full-width half maximum of the Gaussian function) enclosing a cross. The lines are the fitted evolutional tracks of the leading components. The dashed line indicates the primary component (and leading component after mid 2009). The dotted line shows the position of the core.}
\label{evolution_3C111}
\end{figure}

In this work, we focus on the time period 2006 through 2010, which contains 24 MOJAVE epochs with an average image rms noise of 0.2\,mJy\,$\mathrm{beam^{-1}}$ and a maximum of 0.4\,mJy\,$\mathrm{beam^{-1}}$. The restoring beams of different epochs are very similar with an average of (0.84 $\times$ 0.58) mas at P.A. $-8\,^{\circ}$. A peak flux density of 6.54\,Jy was measured in May 2008.
Figure \ref{evolution_3C111} shows the evolution of the parsec scale jet of 3C\,111 observed within the MOJAVE VLBA program at 15\,GHz since May 2008 (excluding six epochs before the ejection of the primary component). A major bright feature appears in May 2008 and is travelling downstream. The source brightness distribution was modelled with the \textsc{CLEAN} algorithm by \cite*{1974A&AS...15..417H} within \textsc{DIFMAP}. 

\subsection{Model Fitting}

\begin{figure*}
\includegraphics[width=0.45\textwidth]{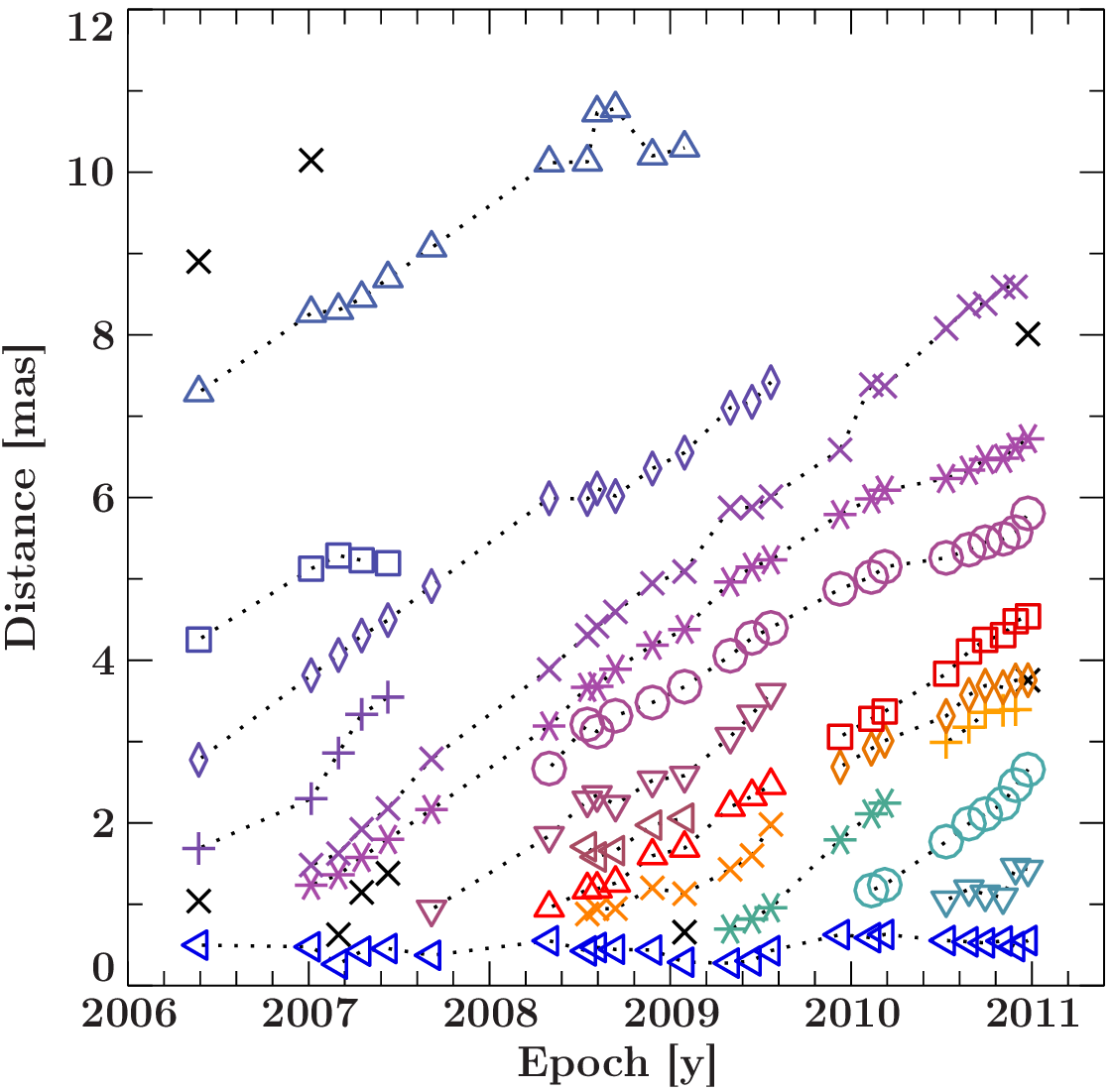}
\includegraphics[width=0.45\textwidth]{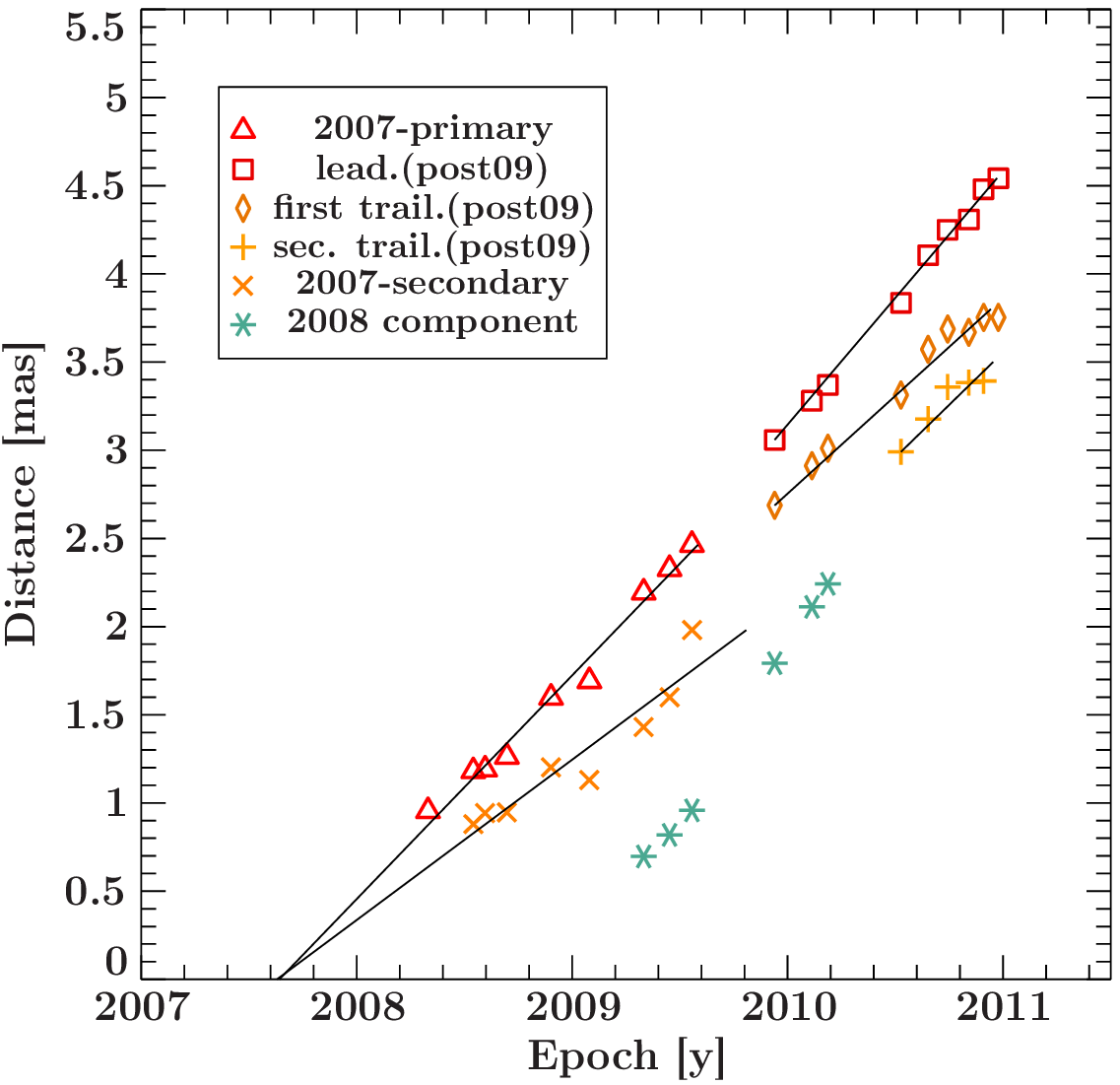}
\caption{Separation of all model-fit components (left) and only components associated with the 2007-outburst (right) with respect to the core as a function of time. Model-fit components which could not be identified in 5 or more epochs are marked with a black cross. The 2007-primary component is presented by triangles, the 2007-secondary component by crosses, the leading component by open rectangles, the first trailing component by open diamonds, the second trailing component by pluses and the 2008-component by stars. Linear regression fits determine the trajectories of the components associated with the 2007-outburst.}
\label{distvsmjd}
\end{figure*}

\begin{figure}
\includegraphics[width=0.45\textwidth]{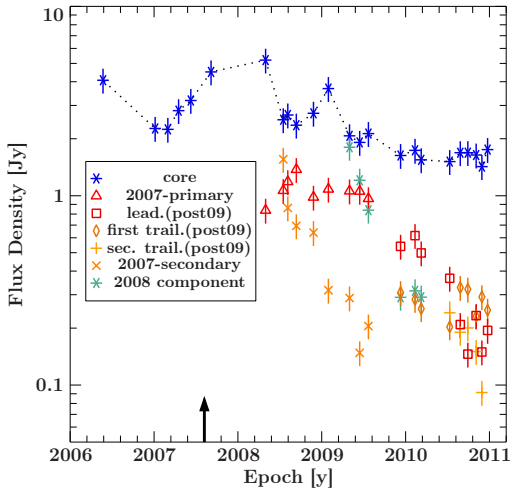}
\caption{Flux density evolution of the``core'' region and the components as a function of time. Conservative errors of 15\% are assumed for the flux densities of the model-fit components. The component symbols are the same as in Fig. \ref{distvsmjd} with the addition of the ``core'' region, presented by stars. The black arrow at the bottom indicates the calculated and almost identical ejection date of the primary and secondary component based on the derived jet kinematics.}
\label{core_x10x11_fluxvsmjd}
\end{figure}

In Figure \ref{distvsmjd}, the radial distances to the stationary component in the core region are plotted for every component as a function of time. The identification throughout the epochs is based on the comparison of the positions and flux densities of these model components. This identification is preliminary and will be discussed in more detail in a forthcoming paper. In the context of this paper we focus on the components which can be associated with the major flux density outburst of 2007 (see Fig. \ref{lc_umrao}).\\
A linear regression fit has been performed to measure component speeds and ejection dates (see Figure \ref{distvsmjd}). The ejection dates of a primary component and a secondary component are quasi identical ($\mathrm{\sim2007.6}$) within the errors and were found to coincide with the peak of the outburst at high frequencies (see Figure \ref{lc_umrao}, right). The components flux density evolution (see Figure \ref{core_x10x11_fluxvsmjd}) shows that the core region was extremely bright during the time of the outburst but dropped significantly after ejecting the bright primary and secondary component. 
The determined apparent speed for the primary component is $\mathrm{3.94\pm0.19\,c}$ and for the secondary component $\mathrm{2.80\pm0.40\,c}$.\\
The primary component remains at a constant flux density of $\mathrm{\sim1\,Jy}$ until mid 2009. After that, the component is splitting into multiple parts: a new leading component with trailing components in its wake.\\
The secondary component has a higher flux density than the primary in the beginning of its lifetime which rapidly decays. This decay suggests that this component disappeared in mid 2009 though an identification with the first trailing component is possible based on position alone.
The calculated apparent speed of the new leading component after mid 2009 is $\mathrm{4.53\pm0.09\,c}$ with the flux density decreasing. The first trailing component has an apparent speed of $\mathrm{3.32\pm0.19\,c}$ and shows a constant flux density evolution. The second trailing component was first observed 2010.53 with a flux density of $\mathrm{\sim240\,mJy}$ and could be modeled until 2010.91 with a flux density of less than 100\,mJy. The flux density evolution of the second trailing component suggests that this component faded away and thus could not be modeled in epoch 2010.98. An association of this second trailing component with the 2008 component is possible based on the flux density evolution and position but needs further investigation of the 2008-outburst.\\
A similiar behaviour with a leading, secondary and trailing components has been seen in the evolution of the components associated with the outburst from 1997 by \cite*{2008ApJ...680..867K}. The components were interpreted as a forward and backward shock with the backward shock fading very fast. In this model, the plasma of the forward shock entered a region of rapidly decreasing external pressure allowing it to expand into the jet ambient medium and accelerate. In the following, the plasma recollimated and trailing features were formed in the wake of the leading component (\cite{2008A&A...489L..29P}).\\

\section{Summary}
In this paper, the ejection of new jet components on parsec scales were associated with a major flux density outburst of 3C\,111 in 2007. It was shown that the major flux density outburst can be associated with the ejection of a primary jet component and secondary component. The evolution of the leading component suggests a split into multiple components. The full multi-epoch kinematical analysis of the VLBA jet of 3C\,111 between 2006 and 2011 will be presented elsewhere (Grossberger et al., in prep).


\end{document}